\documentclass[%
 reprint,
superscriptaddress,
 amsmath,amssymb,
 aps,
]{revtex4-1}

\usepackage{graphicx}% Include figure files
\usepackage{dcolumn}% Align table columns on decimal point
\usepackage{bm}% bold math
\usepackage{hyperref}% add hypertext capabilities
\begin{document}

%\preprint{APS/123-QED}

\title{Single freeform surface design for prescribed input wavefront and target irradiance}% Force line breaks with \\

\author{Christoph B\"osel}
 \email{christoph.boesel@uni-jena.de}

\affiliation{Friedrich-Schiller-Universit\"at Jena, Institute of Applied Physics, Albert-Einstein-Str. 15, 07745 Jena, Germany}%
\author{Herbert Gross}%
\affiliation{Friedrich-Schiller-Universit\"at Jena, Institute of Applied Physics, Albert-Einstein-Str. 15, 07745 Jena, Germany}%
\affiliation{Fraunhofer-Institut f\"ur Angewandte Optik und Feinmechanik, Albert-Einstein-Str. 7, 07745 Jena, Germany}

\begin{abstract}

In beam shaping applications, the minimization of the number of necessary optical elements for the beam shaping process can benefit the compactness of the optical system and reduce its cost. The single freeform surface design for input wavefronts, which are neither planar nor spherical, is therefore of interest.
In this work, the design of single freeform surfaces for a given zero-\'etendue source and complex target irradiances is investigated. Hence, not only collimated input beams or point sources are assumed. Instead, a predefined input ray direction vector field and irradiance distribution on a source plane, which has to be redistributed by a single freeform surface to give the predefined target irradiance, is considered. To solve this design problem, a partial differential equation (PDE) or PDE system, respectively, for the unknown surface and its corresponding ray mapping is derived from energy conservation and the ray-tracing equations.
In contrast to former PDE formulations of the single freeform design problem, the derived PDE of Monge-Amp\`ere type is formulated for general zero-\'etendue sources in cartesian coordinates. The PDE system is  discretized with finite differences and the resulting nonlinear equation system solved by a root-finding algorithm. The basis of the efficient solution of the PDE system builds the introduction of an initial iterate constuction approach for a given input direction vector field, which uses optimal mass transport with a quadratic cost function. After a detailed description of the numerical algorithm, the efficiency of the design method is demonstrated by applying it to several design examples. This includes the redistribution of a collimated input beam beyond the paraxial approximation, the shaping of point source radiation and the shaping of an astigmatic input wavefront into a complex target irradiance distribution.

\begin{description}
\item[PACS numbers]
 \verb+42.15.-i, 42.15.Eq+.
\end{description}
\end{abstract}

\maketitle

%\tableofcontents

\section{Introduction}
\label{sec:1}

The development of numerical design algorithms for the calculation of freeform surfaces for irradiance control in the geometrical optics limit is of great interest in current research.
The task of these methods is thereby the calculation of freeform surfaces, which redistribute a given light source into a required target irradiance pattern. In particular the calculation of \textit{continuous} freeform surfaces, which is considerd here, is of importance for manufacturability and the reduction of possible diffraction effects. Especially the irradiance control with continuous single freeform surfaces for input beams with planar or nonspherical wavefronts lacks of detailed discussion in the literature. Such design algorithms can potentially reduce the number of necessary optical elements or increase the efficiency in practical applications and are investigated in this work.

In principle, the freeform design for shaping of ideal sources can be divided into three major groups. The first group consists of design approaches using Oliker's supporting quadrics method (SQM) \cite{Oliker02_1, Fou10_1, Mic11_1, Oliker17_1}, which is arguably the theoretically best justified, most reliable freeform design method and usable in complex optical systems \cite{Mic11_1}. As it was shown by Oliker \cite{Oliker02_1}, the geometric properties of quadric surfaces can be used effectively for the freeform design process. The shaping of e.g. a point source with a freeform mirror can be done by placing the point source in the common focal point of a number of ellipsoids. By changing the focal parameters of each ellipsoid appropriately, a freeform mirror, which gives the required irradiance distribution on the target area, can be designed by a unification of the ellipsoids. Due to the high computational effort, necessary for high resolution target irradiances and large number of quadrics, the major difficulty, considering the SQM, is the development of efficient algorithms.

The second group of design approaches consists of methods, which are modelling the freeform design problem by a highly nonlinear partial differential equation (PDE) of Monge-Amp\`ere type for point sources \cite{ries02, Wu13_1, Brix15_1} and collimated beams \cite{Wu13_2, Prins14_1}. The main problem is thereby the development of numerical algorithms to solve the PDE. An advantage of the PDE methods over the SQM is the lower computational effort, enabling them to handle high resolution target irradiances. The major drawback of the PDE methods is the lack of theoretical justification and reliability as pointed out by Oliker \cite{Oliker17_1}.

The third group are ray-mapping or source-target mapping methods, respectively \cite{Oli13_1, Brun13_1, Feng13_2, Ma15_1, Boe16_1, Gan17_1}. These are based on the calculation of a ray mapping between the source and the target irradiance distribution an the subsequent construction of the freeform surface from the mapping. The major difficulty of these methods is the finding of an integrable mapping, enabling the design of a freeform surface, which is on the one hand continuous and on the other hand maps the source onto the target distribution. Some of the ray-mapping methods \cite{Brun13_1, Feng13_2, Boe16_1} explore a mapping calculated from optimal mass transport (OMT) theory with a quadratic cost function. This special mapping solves the collimated beam shaping problem in a paraxial approximation \cite{Boe16_1}, but is in general not integrable. As we will demonstrate in this work, the quadratic cost function mapping is nevertheless of great usefulness for the freeform design process for more general optical configurations.

In this work, we want to generalize the design approach for collimated  beam shaping with continuous single freeform surfaces presented in Ref. \cite{Boe16_1} to general input wavefronts of ideal sources, whereby, contrary to former PDE formulations, parallel input beams or point sources are not assumed. Instead, we derive a system of PDE's and a Monge-Amp\`ere equation, respectively, for a predefined input direction vector field and present a detailed numerical algorithm to solve the corresponding problem. As indicated in the previous paragraph, the basis of the numerical algorithm builds thereby a ray mapping, derived from OMT with a quadratic cost function, which will serve as an initial iterate for solving the derived PDE system. As we will demonstrate explicitely, the design algorithm extends the work of Ref. \cite{Boe16_1} firstly to collimated beam shaping in the nonparaxial regime, secondly to the irradiance control of point sources, and thirdly, to other input wavefront shapes.

To introduce the design algorithm, this work is structured as follows. In section \ref{sec:2} and \ref{sec:3}, a PDE system is derived from energy conservation and the ray-tracing equations for single freeform lens and mirror surfaces. For the numerical solution of the PDE system, a method for the calculation of an appropriate initial iterate is introduced in section \ref{sec:4}. In section  \ref{sec:5}, the numerical algorithm for determining the desired freeform surface is presented and in section \ref{sec:6} several design examples, illustrating the effectiveness and usefulness of the design algorithm, are discussed.

\section{Geometry}
\label{sec:2}

\begin{figure}[!htb]
\begin{center}
\begin{tabular}{c}
\includegraphics[width=6cm]{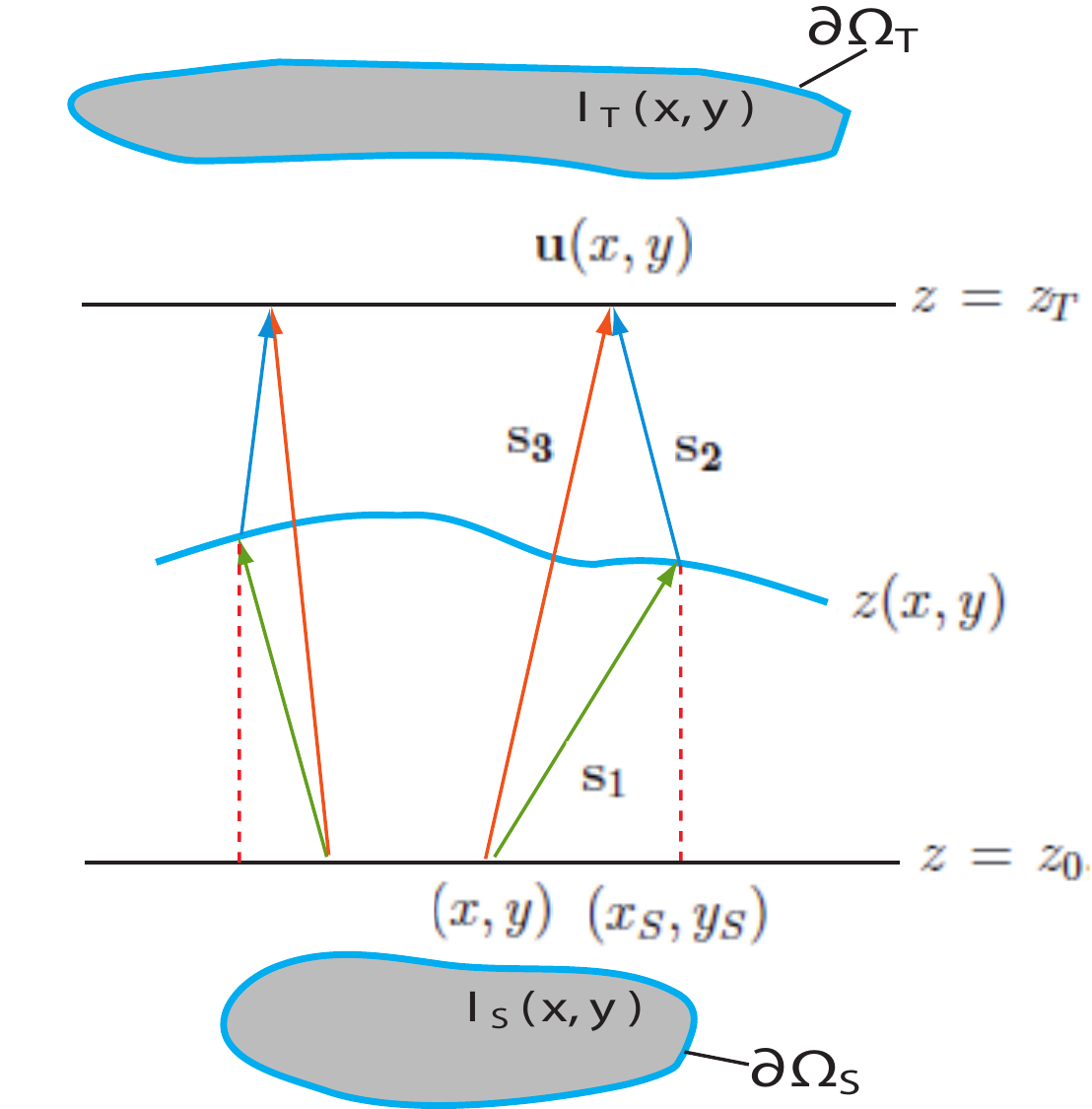}
\end{tabular}
\end{center}
\caption 
{The input $I_S (x,y)$ on the plane $z=z_0$, the output irradiance $I_T (x,y)$ on the plane $z=z_T$ and the input directions $\mathbf{\hat{s}}_1$ are given. The input vector field is redirected by the freeform surface $z(x,y)$ to the target points $\mathbf{u}(x,y)$ according to the law of refraction/reflection to give the desired ouput distribution. The coordinates $(x_S, y_S)$ represent the intersection points of the input vectors with $z(x,y)$.} 
\label{fig:1}
\end{figure}

In the following, the problem of mapping an incoming beam, defined by the ray direction vector field

\begin{gather}\label{eq:1}
\mathbf{\hat{s}}_1 =
\begin{pmatrix}
  (\mathbf{\hat{s}}_1)_x \\
  (\mathbf{\hat{s}}_1)_y \\
  (\mathbf{\hat{s}}_1)_z
\end{pmatrix}, 
\end{gather}

and the irradiance distribution  $I_S (x,y)$ on the plane $z=z_0$, onto a predefined output irradiance distribution $I_T (x,y)$ on the target plane $z=z_T$ is considered (see Fig. \ref{fig:1}).

The goal is thereby the calculation of a continuous reflecting or refracting surface, which enables the mapping of the source onto the target distribution. In a geometrical optics framework, this problem is fully governed by two equations. This is on the one hand the Jacobian equation

\begin{equation}\label{eq:2}
det(\nabla \mathbf{u}(x,y))I_T (\mathbf{u}(x,y))=I_S (x,y),
\end{equation}

describing the relation of the input and output irradiance by the ray mapping or coordinate transformation $\mathbf{u}(x,y)$, respectively, and on the other hand the ray tracing equations for refractive and reflective surfaces

\begin{equation}\label{eq:3}
\begin{aligned}
\mathbf{\hat{s}}_2 =n\mathbf{\hat{s}}_1 + \left\{-n\cdot \mathbf{\hat{n}} \cdot \mathbf{\hat{s}}_1 + \sqrt{1- n^2 [1-(\mathbf{\hat{n}} \cdot \mathbf{\hat{s}}_1)^2]} \right\}\mathbf{\hat{n}},\\
\mathbf{\hat{s}}_2 =\mathbf{\hat{s}}_1 -2(\mathbf{\hat{n}}\cdot \mathbf{\hat{s}}_1)\mathbf{\hat{n}}
\end{aligned}
\end{equation}

with the surface unit normal vector field $\mathbf{\hat{n}}$. Equation (\ref{eq:3}) relates the input coordinates $(x,y)$ and the target coordinates $\mathbf{u}(x,y)$ by the unknown surface $z(x,y)$. Unit vector fields are hereby denoted by the ``hat''. To determine the unknown surface, we want to derive a PDE system for the mapping $\mathbf{u}(x,y)$ and the surface $z(x,y)$.
This can be done by expressing the considered geometry, according to Fig. \ref{fig:1}, through the vector fields 

\begin{equation}\label{eq:4}
\begin{aligned}
\mathbf{s}_1 =
\begin{pmatrix}
  x_S - x \\
  y_S - y \\
  z (x_S, y_S )-z_0
\end{pmatrix}, 
\ \ \
\mathbf{s}_3 =
\begin{pmatrix}
  u_x (x,y) -x\\
  u_y (x,y) -y\\
  z_T - z_0
\end{pmatrix}, 
\\ 
\mathbf{n} =
\begin{pmatrix}
  - \partial_{x_S} z (x_S, y_S ) \\
  - \partial_{y_S} z (x_S, y_S ) \\
  1
\end{pmatrix},
\end{aligned}
\end{equation}

and by calculating the intersection points $\mathbf{u}(x,y)$ with the target plane $z=z_T$ from the relation

\begin{gather}\label{eq:5}
\mathbf{s_1}+ \lambda \cdot \mathbf{\hat{s}_2}= \mathbf{s_3}
\ \ \ \ \
\Rightarrow  
\ \ \ \ \
\lambda =\frac{z_T -z (x_S, y_S )}{(\mathbf{\hat{s}_2})_z}.
\end{gather}

Hereby $x_S, y_S $ and $z (x_S, y_S )$ are the intersection coordinates of the input vector field $\mathbf{s_1}$ with the surface $z (x, y)$. From Eq. (\ref{eq:5}) we then directly get

\begin{equation}\label{eq:6}
\begin{aligned}
u_x (x,y)-x=\lambda \cdot (\mathbf{\hat{s}_2})_x + (\mathbf{s_1})_x,\\
u_y (x,y)-y=\lambda \cdot (\mathbf{\hat{s}_2})_y + (\mathbf{s_1})_y,
\end{aligned}
\end{equation}

which define a relation between the surface $z (x, y)$, the surface gradient and the target coordinates. For the case of collimated input beams, analogues equations were derived by Wu et al. \cite{Wu13_2} and Oliker \cite{Oli14_1}.
Together with Eq. (\ref{eq:2}), Eq. (\ref{eq:6}) builds a PDE system for the unknown functions $u_x (x,y), u_y (x,y)$ and $z (x,y)$.
According to Eq. (\ref{eq:4}), the remaining difficulty is thereby the dependency of Eq. (\ref{eq:6}) on the surface intersection points $(x_S ,y_S )$, which are determined by the surface $z(x,y)$ and therefore unknown. A similar problem arising in their work on the reconstruction of an optical surface from experimental ray data \cite{Rub01_1}, was solved by Rubinstein and Wolansky using the geometrical relation

\begin{equation}\label{eq:7}
\begin{aligned}
\frac{x_S-x}{z(x_S , y_S)-z_0} =\frac{(\mathbf{\hat{s}_1})_x}{(\mathbf{\hat{s}_1})_z} \ \ \ \Rightarrow \ \ \
x_S =\frac{(\mathbf{\hat{s}_1})_x}{(\mathbf{\hat{s}_1})_z}[z(x_S , y_S)-z_0]+x, 
\\
\frac{y_S -y}{z(x_S , y_S)-z_0} =\frac{(\mathbf{\hat{s}_1})_y}{(\mathbf{\hat{s}_1})_z} \ \ \ \Rightarrow \ \ \
y_S =\frac{(\mathbf{\hat{s}_1})_y}{(\mathbf{\hat{s}_1})_z}[z(x_S , y_S)-z_0]+y,
\end{aligned}
\end{equation}

between the given ray directions defined in Eq. (\ref{eq:1}) and the vector field $\mathbf{s}_1$ in Eq. (\ref{eq:4}). By applying this coordinate transformation to the coordinates $x_S$, $y_S$ and the gradient $\nabla_S z(x_S , y_S )= (\partial_{x_S}z(x_S , y_S ), \partial_{y_S} z(x_S , y_S ))$ occuring in Eq. (\ref{eq:4}), Eq. (\ref{eq:6}) can be expressed by the initial coordinates $(x, y)$ on the plane $z=z_0$. After using the chain rule it follows

\begin{equation}\label{eq:8}
\begin{aligned}
\partial_{x_S}z(x_S , y_S )=\frac{(\partial_y y_S) \left[\partial_x z(x_S , y_S)\right] - (\partial_x y_S) [\partial_y z(x_S , y_S)] }{(\partial_x x_S ) \cdot (\partial_y y_S ) - (\partial_x y_S ) \cdot (\partial_y x_S) },
\\
\partial_{y_S}z(x_S , y_S )=\frac{-(\partial_y x_S ) [\partial_x z(x_S , y_S)] + (\partial_x x_S ) [\partial_y z(x_S , y_S)]}{(\partial_x x_S ) \cdot (\partial_y y_S ) - (\partial_x y_S ) \cdot (\partial_y x_S)}.
\end{aligned}
\end{equation}

The remaining derivatives of $x_S$ and $y_S$ can be evaluated with Eq. (\ref{eq:7}).
With these transformations and the definition

\begin{equation}\label{eq:9}
z_S(x,y) \equiv z(x_S, y_S),
\end{equation}

Eq. (\ref{eq:6}) is fully expressed in terms of the $x$-$y$-coordinates. Hence, the Jacobian equation (\ref{eq:2}) and Eq. (\ref{eq:6}) build a system of PDEs for the unknown functions $\mathbf{u}(x,y)$  and $z_S (x,y)$. After solving the equation system for $z_S (x,y)$, the coordinates $(x_S , y_S )$ can than be calculated from Eq. (\ref{eq:7}), which gives the surface $z(x,y)$ on the scattered $x_S$-$y_S$ grid.

\section{PDE system and Monge-Amp\`ere equation}
\label{sec:3}

In the previous section it was shown that the single freeform design problem for zero-\'etendue sources is governed by the PDE system

\begin{equation}\label{eq:10}
\begin{aligned}
\left[(\partial_x u_x) ( \partial_y u_y ) - (\partial_x u_y ) (\partial_y u_x )\right]
I_T (u_x , u_y )=I_S (x,y),
\\
u_x (x,y)-x=[z_T -z_S (x, y)]\frac{(\mathbf{\hat{s}_2})_x}{(\mathbf{\hat{s}_2})_z} + [z_S (x , y)-z_0]\frac{(\mathbf{\hat{s}_1})_x}{(\mathbf{\hat{s}_1})_z} ,
\\
u_y (x,y)-y=[z_T -z_S (x, y)]\frac{(\mathbf{\hat{s}_2})_y}{(\mathbf{\hat{s}_2})_z} + [z_S (x , y)-z_0]\frac{(\mathbf{\hat{s}_1})_y}{(\mathbf{\hat{s}_1})_z} ,
\end{aligned}
\end{equation}

whereby $\mathbf{\hat{s}_1}$ is the predefined input ray direction vector field and $\mathbf{\hat{s}_2}$ is  defined according to section \ref{sec:2} in terms of $z_S(x,y)$, $\partial_x z_S(x,y)$ and $\partial_y z_S(x,y)$.

%Hereby $\mathbf{s}_1$ is given by Eq. (\ref{eq:4}), $\mathbf{\hat{s}_2}$ by the ray-tracing Eq. (\ref{eq:3}), the ray directions  $\mathbf{\hat{s}}_1$ by Eq. (\ref{eq:1}) and the surface normal vector field $\mathbf{n}$ by plugging the Eqs. (\ref{eq:8}) and (\ref{eq:7}) into Eq. (\ref{eq:4}).

For a given source and target distribution $I_S(x,y)$ and $I_T(x,y)$, and a given input ray direction vector field $\mathbf{\hat{s}_1}$ the equation system in Eq. (\ref{eq:10}) has to be solved simultaneously for the surface $z_S (x,y)$ and the ray mapping $\mathbf{u}(x,y)$.

Equation (\ref{eq:10}) can be further simplified by plugging $u_x (x,y)$ and $u_y (x,y)$ into the Jacobian equation to obtain a nonlinear second order PDE of Monge-Amp\`ere type for the function $z_S (x,y)$:

\begin{equation}\label{eq:11}
\begin{aligned}
A\partial_{xx} z_S +2B \partial_{xy} z_S + C\partial_{yy} z_S 
+\left[(\partial_{xx} z_S) (\partial_{yy} z_S)  -(\partial_{xy} z_S )^2 \right]=D.
\end{aligned}
\end{equation}

Hereby, $A, B, C, D$ are functions  of $x, y, z_S(x,y), \partial_x z_S(x,y) $ and $\partial_y z_S(x,y) $, which can be calculated straightforeward.
\\
Compared to the Eq. system (\ref{eq:10}), the Eq. (\ref{eq:11}) has numerically the advantage of a lesser number of design variables. Despite this advantage, the numerical method presented in the following will nevertheless be based on the description in Eq. (\ref{eq:10}). In contrast to the Monge-Amp\`ere equation in Eq. (\ref{eq:11}), we will therefore omit the discretization of second order derivatives. In addition to that, the boundary conditions can be directly controlled through the mapping $\mathbf{u}(x,y)$ by requiring that boundary points of the support of $I_S(x,y)$ are mapped onto boundary points of the support of $I_T(x,y)$. This is contrary to the Monge-Amp\`ere equation for which the same goal has to be accomplished through the surface gradients $\nabla z_S (x,y)$ at the boundary of the surface, which leads to the dependency of the boundary conditions on the input field $\mathbf{\hat{s}}_1$ . Furthermore, the properties of the Jacobian equation and the mapping equations in Eq. (\ref{eq:10}) can be controlled separately.

\section{Initial iterate}
\label{sec:4}

The construction of freeform surfaces for complex target irradiances from nonlinear PDEs, like Eqs. (\ref{eq:10}) or (\ref{eq:11}) is usually done by discretizatizing the PDEs and solving the resulting nonlinear equation system by an iterative process. A high resolution target irradiance distribution and therefore large number of unknowns requires in general an appropriate initial iterate to ensure (fast) convergence. Especially for discretization schemes, which are not provably convergent, those initial iterates are of great importance. For the published PDE methods for single freeform surface design an inital iterate producing a uniform illumation on the target plane or other simple surface shapes are usually chosen.
\\
Here, we present a different method to construct an initial iterate for a given input direction vector field $\mathbf{\hat{s}_1}$. It is based on the OMT with the quadratic cost function and inspired by Ref. \cite{Rub01_1}. Following the argumentation from Ref. \cite{Rub01_1}, an unknown surface can be experimentally reconstructed from the known input ray directions and measured target plane intersection points of the rays send out from $(x,y)$.
Since the target points are not given for the freeform design problem considered here, the approach from Ref. \cite{Rub01_1} can not be directly applied. Nevertheless, as we will demonstrate in the following, the approach is of great usefullness to generate an initial iterate for the freeform surface design.

\subsection{Initial mapping}
\label{sec:4.1}

For the single freeform surface design process a frequently used mapping $\mathbf{u}(x,y)$ is calculated from OMT by considering Eq. (\ref{eq:2}) together with the quadratic cost function condition

\begin{equation}\label{eq:12}
d(I_S, I_T)^2 = \inf_{\mathbf{u}\in M} \int |\mathbf{u}(\mathbf{x})-\mathbf{x}|^2 I_S (\mathbf{x}) d\mathbf{x},
\end{equation}

which we will refer to in the following by $\mathbf{u^{\infty}}(x,y)$. $M$ in Eq. (\ref{eq:12}) denotes the set of energy preserving mappings. This is arguably the most studied OMT problem in numerical mathematics and provides us therefore with efficient algorithms for the mapping calculation.
As it was demonstrated in several publications, the corresponding mapping $\mathbf{u^{\infty}}(x,y)$ in general will not generate a continuous surface $z(x , y )$ fulfilling the required design goal of mapping the given input $I_S (x,y)$ onto the desired output irradiance $I_T (x,y)$. However, due to its properties, the mapping $\mathbf{u^{\infty}}(x,y)$ appears to be a good candidate for creating an appropriate initial iterate. According to Eq. (\ref{eq:12}), the quadratic cost function can in principle be thought of as describing a ray propagation in a homogeneous medium \cite{Rub04_1}. Since a continuous surface can be interpreted as a perturbation of this free propagation, it can be expected that the mapping $\mathbf{u^{\infty}}(x,y)$ serves as a reasonable initial iterate for solving Eq. (\ref{eq:10}). From a numerical point of view it is also helpful that $\mathbf{u^{\infty}}(x,y)$ already fulfills the Jacobian equation in Eq. (\ref{eq:10}), which therefore basically constrains all numerical corrections to $\mathbf{u^{\infty}}(x,y)$ to the space of energy preserving mappings.
Hence, we will use the mapping defined by Eqs. (\ref{eq:2}) and (\ref{eq:12}) as an initial mapping for solving Eq. (\ref{eq:10}). As in the previous publications \cite{Boe16_1, Boe17_1}, we will make use of the algorithm by Sulman et al. \cite{Sul11_1} for the examples in section \ref{sec:6} for demonstrational purposes.

\subsection{Initial surface}
\label{sec:4.2}

After its calculation, the mapping $\mathbf{u}^{\infty}(x,y)$ can be used to construct the initial surface $z_S^{\infty}(x , y)$ as demonstrated in the following. The resulting initial surface construction equations are thereby a generalization of the surface construction equations used in Ref. \cite{Boe16_1} to noncollimated beams.

The derivation can be done by using the law of refraction/reflection

\begin{equation}\label{eq:13}
\frac{\mathbf{n}(x_S , y_S)}{(\mathbf{n}(x_S , y_S))_z}=\frac{n_1 \mathbf{\hat{s}}_1 -n_2 \mathbf{\hat{s}}_2}{(n_1 \mathbf{\hat{s}}_1 -n_2 \mathbf{\hat{s}}_2)_z}=\nabla_S (z-z_S(x , y ))
\end{equation}

and applying again the coordinate transformation defined in Eq. (\ref{eq:7}) to invert the gradient by using

\begin{equation}\label{eq:14}
\begin{aligned}
\partial_{x_S}z_S^{\infty}(x , y )
=
-\frac{n_1 (\mathbf{\hat{s}}_1 )_x -n_2 (\mathbf{\hat{s}}_2  )_x}{n_1 (\mathbf{\hat{s}}_1)_z -n_2 (\mathbf{\hat{s}}_2 )_z}
\equiv
 n_x (x,y),
\\
\partial_{y_S}z_S^{\infty}(x , y )
=
-\frac{n_1 (\mathbf{\hat{s}}_1 )_y -n_2 (\mathbf{\hat{s}}_2  )_y}{n_1 (\mathbf{\hat{s}}_1)_z -n_2 (\mathbf{\hat{s}}_2 )_z}
\equiv
 n_y (x,y).
\end{aligned}
\end{equation}

After using the chain rule, this gives the differential equations \cite{Rub01_1}

\begin{equation}\label{eq:15}
\begin{aligned}
\partial_{x}z_S^{\infty}(x , y )
=
\\
\frac{n_x (x,y)+\left[n_x (x,y)\partial_x \left(\frac{(\mathbf{\hat{s}}_1 )_x}{(\mathbf{\hat{s}}_1 )_z }\right)+n_y (x,y)\partial_x \left(\frac{(\mathbf{\hat{s}}_1 )_y}{(\mathbf{\hat{s}}_1 )_z }\right) \right](z_S^{\infty}(x,  y )-z_0) }{1-n_x (x,y)\frac{(\mathbf{\hat{s}}_1 )_x}{(\mathbf{\hat{s}}_1 )_z }- n_y (x,y)\frac{(\mathbf{\hat{s}}_1 )_y}{(\mathbf{\hat{s}}_1 )_z } },
\\
\partial_{y}z_S^{\infty}(x , y )
=
\\
\frac{n_y (x,y)+\left[n_x (x,y)\partial_y \left(\frac{(\mathbf{\hat{s}}_1 )_x}{(\mathbf{\hat{s}}_1 )_z }\right)+n_y (x,y)\partial_y \left(\frac{(\mathbf{\hat{s}}_1 )_y}{(\mathbf{\hat{s}}_1 )_z }\right) \right](z_S^{\infty}(x,  y )-z_0) }{1-n_x (x,y)\frac{(\mathbf{\hat{s}}_1 )_x}{(\mathbf{\hat{s}}_1 )_z }- n_y (x,y)\frac{(\mathbf{\hat{s}}_1 )_y}{(\mathbf{\hat{s}}_1 )_z } }
\end{aligned}
\end{equation}

for the initial surface $z_S^{\infty}(x , y )$. For a given input direction vector field, defined by Eq. (\ref{eq:1}), and given target points $\mathbf{u}^{\infty}(x,y)$, Eq. (\ref{eq:15}) can be integrated along an arbitrary path on the support of $I_S (x,y)$ to generate the initial surface $z_S^{\infty}(x , y )$.

Together with the initial mapping $\mathbf{u}^{\infty}(x,y)$, the surface $z_S^{\infty}(x , y )$ builds an initial iterate for solving the PDE system (\ref{eq:10}).

We want to remark that the initial surface construction approach, presented in this section can also be used for the Monge-Amp\`ere equation in Eq. (\ref{eq:11}). To implement the boundary conditions in that case, the surface boundary normal vectors defined by Eq. (\ref{eq:15}) have to be chosen in a way that the boundary of $I_S (x,y)$ is mapped onto the boundary of $I_T (x,y)$.

\section{Numerics}
\label{sec:5} 

\subsection{Discretization}
\label{sec:5.1} 

After the calculation of the initial iterate, the PDE system (\ref{eq:10}) has to be solved. Hence, we have to discretize (\ref{eq:10}) and solve the resulting nonlinear equation system by a root-finding algorithm. To do that, we follow Ref. \cite{Boe17_1} by introducing the corrections $\Delta \mathbf{u}(x,y)$ and $\Delta z_S(x,y)$ to $\mathbf{u}^{\infty} (x,y)$ and $z^{\infty} (x,y)$:

\begin{equation}\label{eq:16}
\begin{aligned}
\mathbf{u}(x,y) &=& \mathbf{u}^{\infty} (x,y)+\Delta \mathbf{u}(x,y),\\
z_S(x,y) &=& z_S^{\infty} (x,y)+\Delta z_S(x,y)
\end{aligned}
\end{equation}

and redefine the initial iterates for the solving process to be $\Delta \mathbf{u}(x,y)=0$ and $\Delta z(x,y)=0$.
Accordingly, Eq. (\ref{eq:10}) becomes a PDE system for $\Delta \mathbf{u}(x,y)$ and $\Delta z_S(x,y)$. To solve the PDE system, we discretize the functions $\Delta u_x (x,y), \Delta u_y (x,y), \Delta z_S(x,y)$ and their derivatives in Eq. (\ref{eq:10}) by central finite differences

\begin{equation}\label{eq:17}
\begin{aligned}
&\partial_x ( \Delta z_S) 
\rightarrow 
\frac{1}{2\Delta x } [(\Delta z_S)_{i;j+1} - (\Delta z_S)_{i;j-1}],
\\
&\partial_y ( \Delta z_S)
\rightarrow
\frac{1}{2\Delta y } [(\Delta z_S)_{i+1;j} - (\Delta z_S)_{i-1;j}]
\end{aligned}
\end{equation}

for the inner points $i; j =2,...,N-1$ and second order finite differences

\begin{equation}\label{eq:18}
\begin{aligned}
\partial_x ( \Delta z_S)
\rightarrow
-\frac{1}{2\Delta x } [3(\Delta z_S)_{i;j+2} - 4(\Delta z_S)_{i;j+1}+(\Delta z_S)_{i;j}],
\\
\partial_y ( \Delta z_S)
\rightarrow
-\frac{1}{2\Delta y } [3(\Delta z_S)_{i+2;j} - 4(\Delta z_S)_{i+1;j}+(\Delta z_S )_{i;j}],
\end{aligned}
\end{equation}

for the boundary points. Additionally, we apply boundary conditions by requiring that the boundary points of the source distribution $I_S (x,y)$ are mapped onto the boundary points of the target distribution $I_T (x,y)$. For the mapping of square supports, which is demonstrated in section \ref{eq:6}, the boundary condition reads therefore:

\begin{equation}\label{eq:19}
\begin{aligned}
\Delta u_x(-0.5,y)=\Delta u_x(0.5,y) = 0, \ \ \ y\in [-0.5,0.5],\\
\Delta u_y(x,-0.5)=\Delta u_y(x,0.5) = 0, \ \ \ x\in [-0.5,0.5].
\end{aligned}
\end{equation}

After the discretization of Eq. (\ref{eq:10}), the resulting nonlinear Eq. system can then be solved by a root-find algorithm for the $3\cdot N^2$ unknown values $(\Delta u_x)_{i;j}, (\Delta u_y)_{i;j}$ and $(\Delta z_S )_{i;j}$. The solution will lead to a discretized surface $z_S(x,y)$ from which, according to Eq. (\ref{eq:7}), the scattered grid points $(x_S , y_S )$ can be calculated. As in Ref. \cite{Boe17_1}, we use MATLAB's \textit{fsolve()} function with the trust-region-reflective solver for the solution of the nonlinear Eq. system (\ref{eq:10}).

\subsection{Design algorithm}
\label{sec:5.2} 

The design algorithm developed in the previous sections can be summarized as follows:

\begin{enumerate}
	\item Define the input ray direction vector field $\mathbf{\hat{s}_1}$ and the input irradiance $I_S (x,y)$ on $z=z_0$ and the output irradiance $I_T (x,y)$ on $z=z_T$.
	\item Calculate the initial map $\mathbf{u}^{\infty}(x,y)$ between 
			$I_S (x,y)$ and $I_T (x,y)$.
	\item Integrate Eq. (\ref{eq:15}) along an arbitrary path on the support of $I_S(x,y)$ to get the initial surface $z_S^{\infty}(x,y)$.
	\item Discretize the PDE system in Eq. (\ref{eq:10}) to get a nonlinear equation system.
	\item Apply the boundary conditions and solve the nonlinear equation system with $\mathbf{u}^{\infty}(x,y)$ and $z_S^{\infty}(x,y)$ as the initial iterate to get $\mathbf{u}(x,y)$ and $z_S(x,y)$. 
\end{enumerate}

As mentioned above, for a general input vector field $\mathbf{\hat{s}_1}$, the presented design algorithm gives us the freeform surface $z(x,y)$ on the scattered grid points $(x_S , y_S )$. The quality of the surface $z_S(x,y)$ is thereby already characterized by Eq. (\ref{eq:10}), which is equivalent to a ray-tracing calculation. For manufacturing processes it might still be useful to evaluate the surface quality in a ray-tracing simulation. We therefore interpolate the scattered data points $z(x_S, y_S)$ onto a regular grid and import it to a ray-tracing software. The interpolation can be done by scattered data interpolation with e.g. MATLAB 2015b's \textit{griddata()} function. However, in our experience this function has two major drawbacks for the interpolation of freeform surfaces. This is its inability (for large data sets) to extrapolate outside of the convex hull of the grid points, which leads to inaccuracies at the boundary of the freeform surface in form of zig-zag patterns at the target irradiance boundary. Furthermore, the interpolation with \textit{griddata()} leads to oscillations in the freeform surface, depending on the stepsize of the interpolation grid and the interpolation method.

To overcome this limitation, we therefore use the \textit{RegularizeData3D()} function from the MATLAB file exchange website \cite{MatFE_1}. This function provides the possibility to choose how smooth the interpolated data is fitted to the scattered data. In our examples, we made good experience with a small smootheness factor for low noise data. Additionally, \textit{RegularizeData3D()} extrapolates outside the convex hull of the scattered data grid, which may also lead to slight inaccuracies at the boundary of the freeform surface, which lower the energy efficiency, as we will point out in the example section \ref{sec:6}.

Since the initial mapping $\mathbf{u}^{\infty}(x,y)$ defined by Eqs. (\ref{eq:2}) and (\ref{eq:12}) is in general not integrable, the integration of Eq. (\ref{eq:15}) will be path dependent and therefore lead to different initial surfaces $z_S^{\infty}(x,y)$ for different integration paths. Nevertheless, in our experience the path dependency has no significant influence on the suitability for using $z_S^{\infty}(x,y)$ as an initial iterate.

\section{Examples}
\label{sec:6} 

In this section, we want to demonstrate basically three things. Firstly, the design algorithm presented in section \ref{sec:5} extends the work of Ref. \cite{Boe16_1} on collimated beam shaping with single freeform surfaces to the nonparaxial regime. Secondly, the design algorithm can be applied to the design of single freeform surfaces for shaping of point light sources, which is frequently discussed in literature. And thirdly, we want to demonstrate the applicability to a nonstandard design example by dealing with an input wavefront, which is neither planar nor spherical. This is done in the following three subsections. All the examples in this sections were calculated on an Intel Core i3 at $2 \times 2.4$Ghz with $16$GB RAM. To make the ray-tracing results comparable, we used for each of the examples the target distribution ``lake" (see Fig. \ref{fig:3}(b)) with a resolution of $250 \times 250$ pixels. For the calculation of the initial surface by Eq. (\ref{eq:15}), we used MATLAB's \textit{ode45} solver with tolerances of $10^{-10}$, which is in our experience necessary for a legitimate comparison of the initial and final surface in a ray-tracing simulation. For the root finding of Eq. (\ref{eq:10}) by \textit{fsolve()} tolerances of $10^{-3}$ were used. For the examples with lenses refractive indices of $n=1.5$ were used. The ray-tracing simulations for the following examples are done with $200 \cdot 10^6$ rays and detectors with $250 \times 250$ pixels by using a self-programmed MATLAB ray-tracing toolbox.
Since the design examples are calculated in a geometrical optics framework, we will use arbitrary units as length units and omit the explicit declaration of lengths by ``a.u.".

\subsection{Collimated beam shaping beyond the paraxial approximation}
\label{sec:6.1} 

In Ref. \cite{Boe16_1} it was shown that the single freeform surface design problem for collimated input beams

\begin{gather}\label{eq:20}
\mathbf{\hat{s}_1} =
\begin{pmatrix}
  0 \\
  0 \\
  1
\end{pmatrix}
\end{gather}

can be solved by the quadratic cost function ray mapping, defined by Eqs. (\ref{eq:2}) and (\ref{eq:12}), in the paraxial regime. Here, we demonstrate that the presented algorithm from section \ref{sec:5} directly extends the results from Ref. \cite{Boe16_1} to nonparaxial calculations. Since the quadratic cost function mapping solves the single freeform surface design problem exactly for infinite distances between the freeform surface and the target plane\cite{Boe16_1}, it can be argued analogously to Ref. \cite{Boe17_1} that the numerical convergence of the algorithm can theoretically be ensured by a stepwise reduction of the target plane distance $z_T$. However, in our experience, this stepwise reduction is in a practical calculation not necessary to ensure numerical convergence.

\begin{figure}[!htb]
\begin{center}
\begin{tabular}{c}
\includegraphics[width=\linewidth]{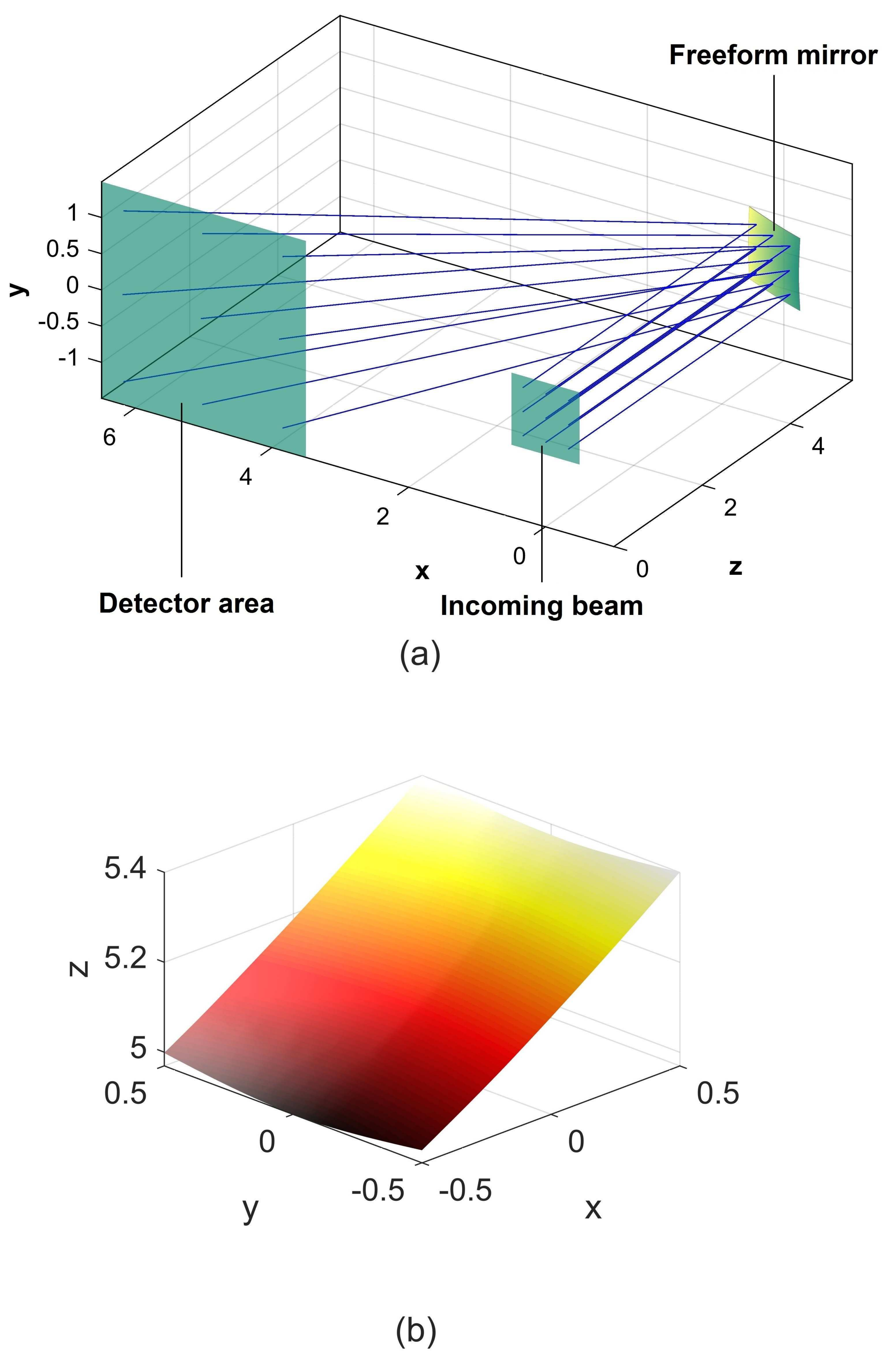}
\end{tabular}
\end{center}
\caption 
{(a) System layout. The incoming collimated Gaussian beam is reflected by the freeform mirror to give the required target distribution ``lake" at the predefined position. (b) Freeform mirror.}
\label{fig:2}
\end{figure}

From an application point of view, the extension to the nonparaxial regime is especially important for optical configurations in which the distance between the target plane and the freeform surface is small, the target area is strongly shifted or scaled, or if there are caustics between the target plane and the freeform surface. In the following, we therefore demonstrate the utility of the presented algorithm by designing a freeform mirror, which shifts and scales the target area compared to the position and size of the input beam. The corresponding system layout, considered in this section, can be seen in Fig. \ref{fig:2}, showing a target area with a side length, which is $3$ times as large as the input beams side length of $1$ and shifted by $5$. The distance between the freeform mirror and the target plane $z_T$ is approximately $5$.

\begin{figure}[!htb]
\begin{center}
\begin{tabular}{c}
\includegraphics[width=\linewidth]{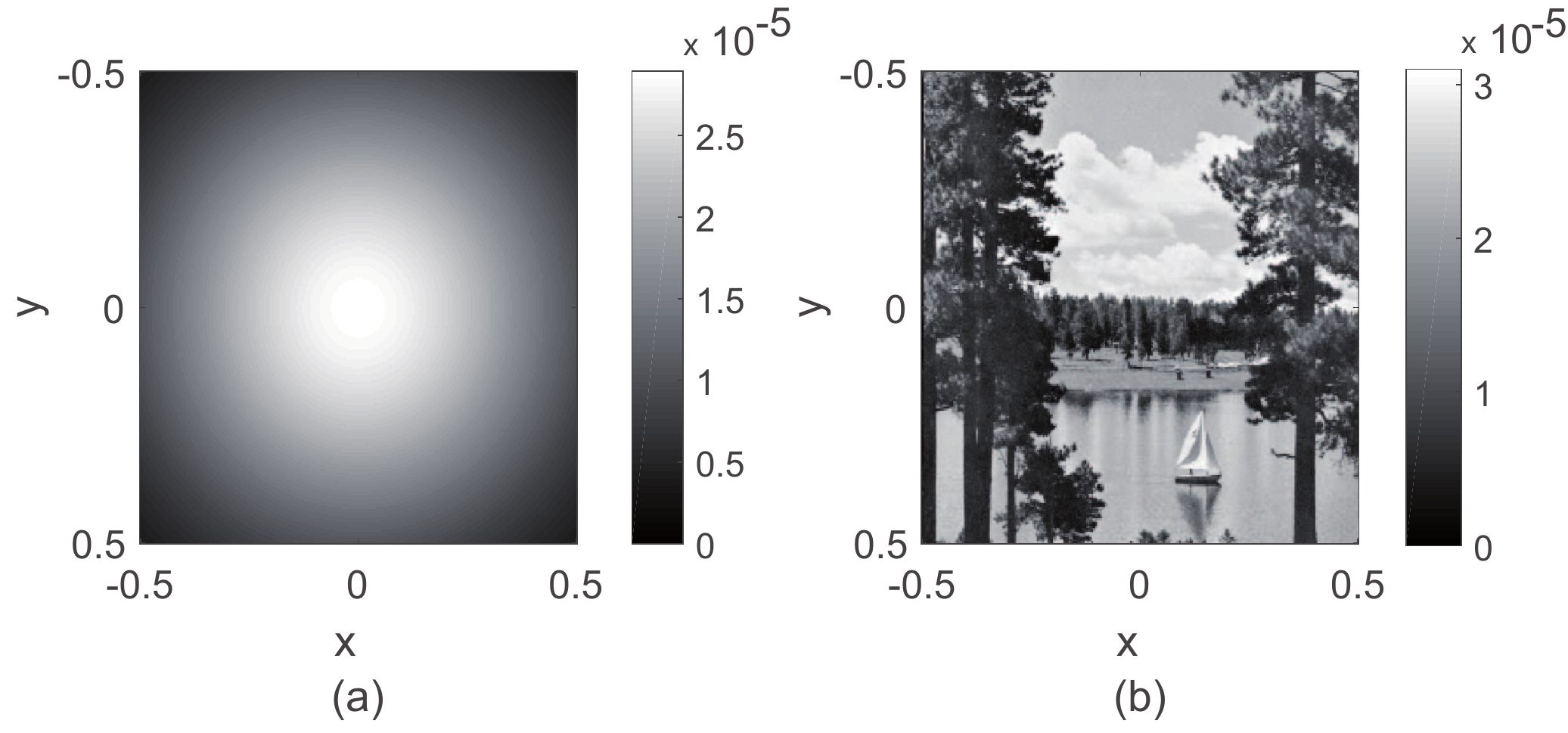}
\end{tabular}
\end{center}
\caption 
{(a) Normalized Gaussian input irradiance distribution $I_S(x, y)$ and (b) normalized output irradiance distribution $I_T(x,y)$ ``lake". The output irradiance in (b) has the side length of $1$, but will be scaled and shifted for comparison with the ray tracing simulations according to the layouts of each example. $I_S(x, y)$ and the input ray direction vector field in Eq. (\ref{eq:20}) at $z=z_0$ and $I_T(x,y)$ at $z=z_T$  build the input of the design algorithm presented in section \ref{sec:5}.}
\label{fig:3}
\end{figure}

\begin{figure}[!htb]
\begin{center}
\begin{tabular}{c}
\includegraphics[width=\linewidth]{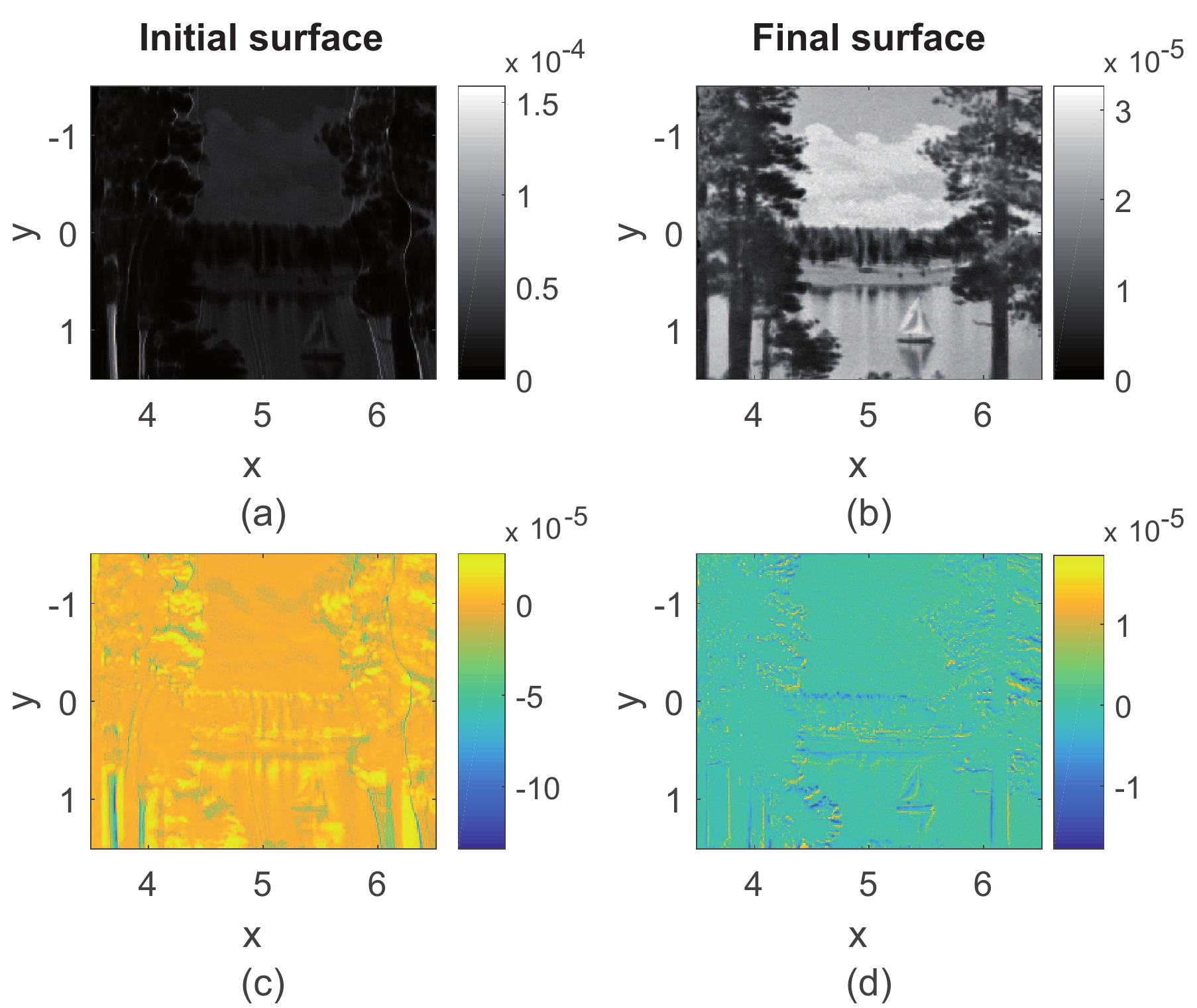}
\end{tabular}
\end{center}
\caption 
{Target irradiance distribution $I_T(x,y)$ from the ray tracing simulation produced by (a) the initial surface and (b) the optimized surface. Difference between the predefined distribution $I_T(x,y)$ and the ray tracing simulation $I_{T,RT}(x,y)$ for (c) the initial surface and (d) the optimized surface. According to (c) the illumation pattern from the initial surface is strongly distorted, whereas the optimized surface produces only slight deviations from the predefined pattern.}
\label{fig:4}
\end{figure}

For the irradiance distribution, we choose the Gaussian distribution, which is represented in Fig. \ref{fig:3} (a)
and the test image ``lake" as the required target distribution (Fig. \ref{fig:3} (b)). According to section \ref{sec:5}, we first calculate the initial mapping between the input and output irradiances ($320 sec$), then the initial surface from Eq. (\ref{eq:15}) ($68 sec$) and finally solve the PDE system in Eq. (\ref{eq:10}) ($208 sec$).
After the construction of the initial and final freeform mirror, we evaluate their quality by a ray-tracing simulation. The corresponding results and comparisons with the algorithm presented in Ref. \cite{Boe16_1} , which is equal to the initial iterate of the root finding, are shown in Fig. \ref{fig:4}. Additionally, the $rms$ and the peak-to-valley of $\Delta I_T (x,y)=I_{T}(x,y)-I_{T,RT}(x,y) $, and the correlation coefficient $corr_{I_T}$ between the predefined irradiance $I_{T}(x,y)$ and the irradiance from the ray tracing $I_{T,RT}(x,y)$, and the energy efficiency $\eta$ are given in Table \ref{table:1} . The results thereby show clearly the improvement of the presented design algorithm compared to Ref. \cite{Boe16_1} and the mapping from the quadratic cost function (\ref{eq:12}), respectively.

\begin{table}[!htb]
\centering
\caption{\bf Comparison of $\Delta I_T$ for example ``lake" with a collimated input beam.}
\begin{tabular}{ccc}
\hline
	 & $z^{\infty}(x,y)$ & $z(x,y)$ \\
\hline
$rms_{\Delta I_T}$	& $9.2288\cdot 10^{-6}$	& $2.3472\cdot 10^{-6}$	\\
$pv_{\Delta I_T}$	& $1.6158\cdot 10^{-4}$	& $3.6349\cdot 10^{-5}$	\\
$corr_{I_T}$		& $0.5750$				& $0.9607$				\\
$\eta$				& $85.66 \% $			& $99.98 \% $		\\
\hline
\end{tabular}
  \label{table:1}
\end{table}

\subsection{Point source}
\label{sec:6.2} 

Besides the collimated beam shaping with a single freeform surface, we want to apply the design algorithm to the other standard configuration considered in literature, which is the mapping of a point source onto a required irradiance distribution on a target plane $z_T$. In the case of single freeform lens design, the point source is assumed to be in the medium or the rays emitted by the point source first hit a spherical surface, respectively. The ray directions are therefore first unaltered and then redirected by the freeform surface to give the predefined target distribution $I_T (x,y)$.

\begin{figure}[!htb]
\begin{center}
\begin{tabular}{c}
\includegraphics[width=\linewidth]{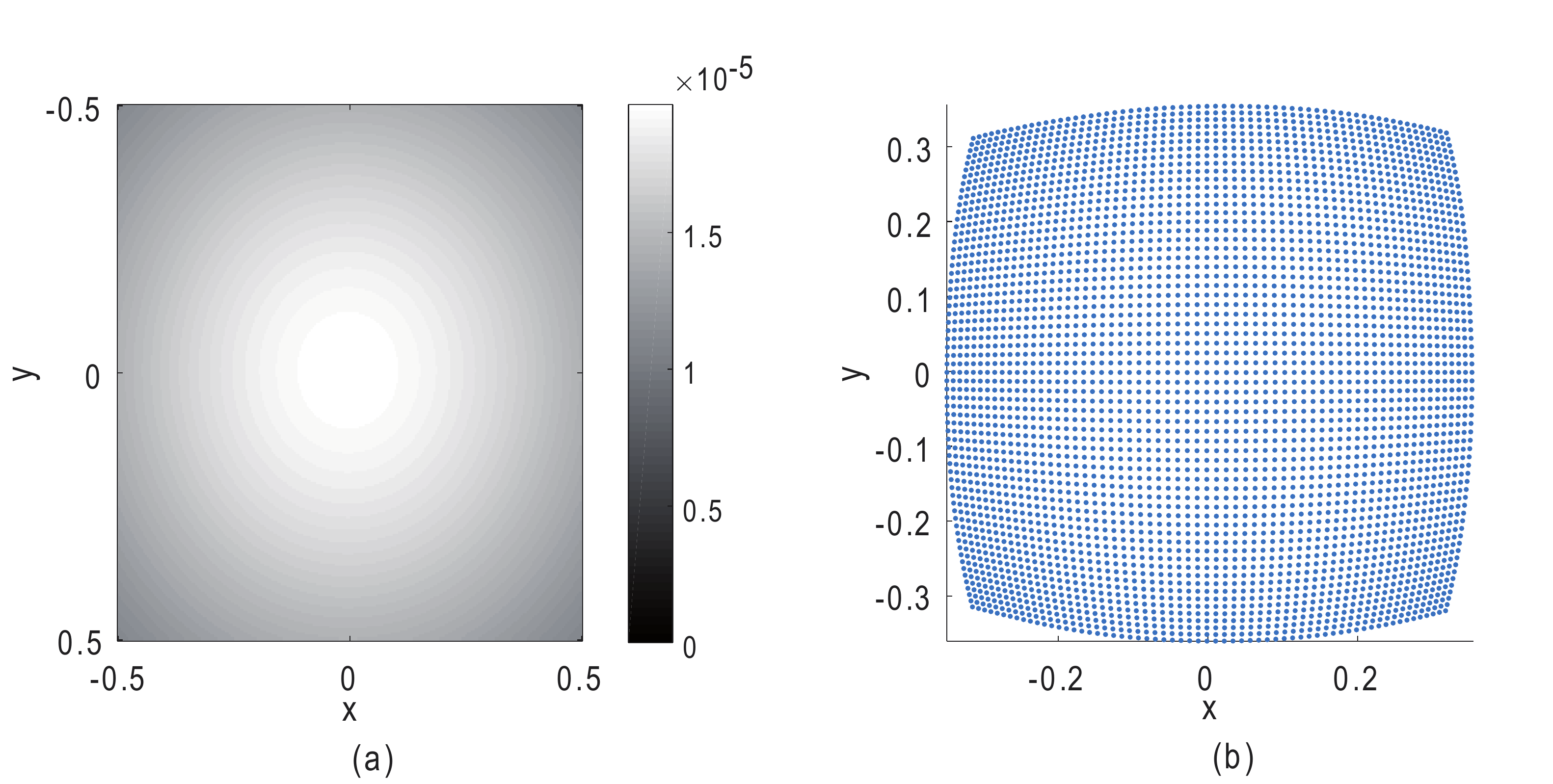}
\end{tabular}
\end{center}
\caption 
{(a) Projection of the lambertian source onto the input plane $z=z_0$, which serves as the irradiance distribution $I_S (x,y)$ for the calculation of $\mathbf{u}^{\infty}(x,y)$ by OMT and together with the input ray direction vector field in Eq. (\ref{eq:21}) and $I_T(x,y)$ as the input of the design algorithm. (b) After solving the PDE system in Eq. (\ref{eq:10}) for the surface $z_S (x,y)$, the scattered grid points $(x_S , y_S)$ can be calculated from Eq. (\ref{eq:7}), which gives $z (x_S ,y_S )$. This function is then interpolated and extrapolated onto a regular grid.}
\label{fig:6}
\end{figure}

Since we want to apply the algorithm from section \ref{sec:5}, we first have to project the point source onto the input plane $z=z_0$, giving the distribution $I_S(x,y)$. Subsequently, the mapping $\mathbf{u}^{\infty}(x,y)$ between the projected distribution $I_S (x,y)$ on $z=z_0$ and the target distribution $I_T (x,y)$ on $z=z_T$ can be calculated. For the point source, we choose a lambertian distribution with a maximum opening angle of $30$ degrees, which refers to the angle between the $z$-axis and the corner points of the quadratic aperture and leads to the irradiance distribution $I_S(x,y)$ presented in Fig. \ref{fig:6}(a).

The input vector field $\mathbf{\hat{s}_1}$ on $z=z_0$ is given by 

\begin{gather}\label{eq:21}
\mathbf{\hat{s}_1} =
\frac{1}{\sqrt{x^2+y^2+z_0^2}}
\begin{pmatrix}
  x \\
  y \\
  z_0
\end{pmatrix}.
\end{gather}

\begin{figure}[!htb]
\begin{center}
\begin{tabular}{c}
\includegraphics[width=\linewidth]{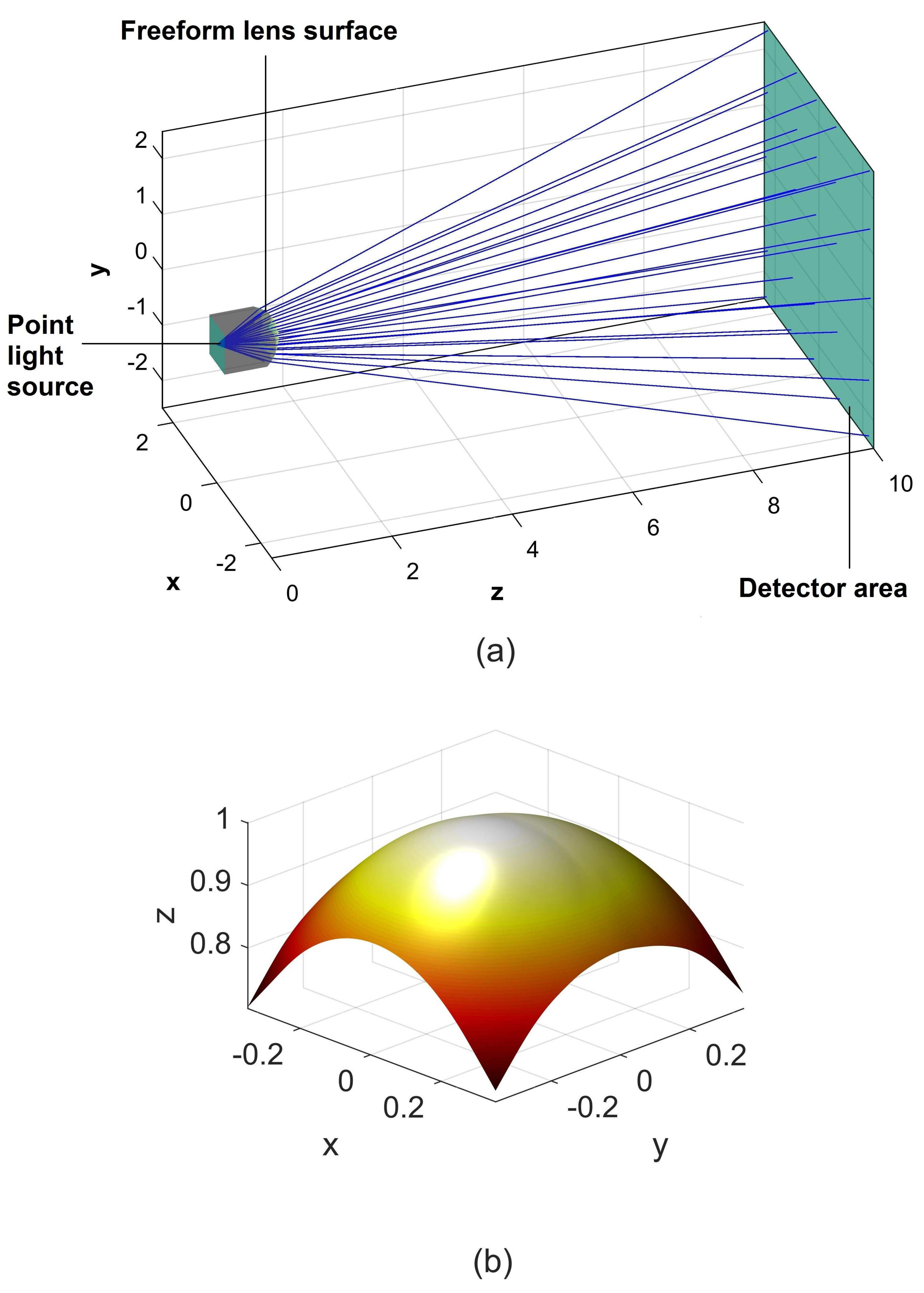}
\end{tabular}
\end{center}
\caption 
{(a) System Layout. The point source at $z=0$ with a lambertian intensity profile is redistributed by the freeform lens to give the required target irradiance at $z=10 $. (b) Freeform lens surface. Since \textit{RegularizeData3D()} extrapolates the surface to a square area, the corner areas are outside of the maximum opening angle of $30$ degrees and therefore irrelevant and not considered in the ray-tracing simulation.}
\label{fig:5}
\end{figure}

The placement of the plane $z=z_0$ relative to the point source does not matter. Depending on the value of $z_0$ only the mapping $\mathbf{u}^{\infty}(x,y)$ has to be scaled by a constant factor to create the predefined target distribution area size, which we choose to have a side length of $5$. For the distance between the point source and the freeform surface we choose $z(0,0)=1$, which enters the design process as an integration constant in Eq. (\ref{eq:15}), and for the distance between the point source and the target plane, we choose $z_T=10$. The corresponding system layout can be seen in Fig. \ref{fig:5}.

The calculation of the initial mapping $\mathbf{u}^{\infty}(x,y)$ took approximately $378 sec$, the initial surface $z_S^{\infty}(x,y)$ $183 sec$ and the final surface $z_S(x,y)$ about $213 sec$. After the calculation of the freeform surface $z_S (x,y)$ by the presented design algorithm an additional complication, compared to the collimated beam shaping, arises since the freeform surface is calculated on the nonuniform $x_S$-$y_S$-grid. Fig. \ref{fig:6}(b) shows every 4th point of the scattered grid points. The outermost points define the aperture of the freeform lens.

\begin{figure}[!htb]
\begin{center}
\begin{tabular}{c}
\includegraphics[width=\linewidth]{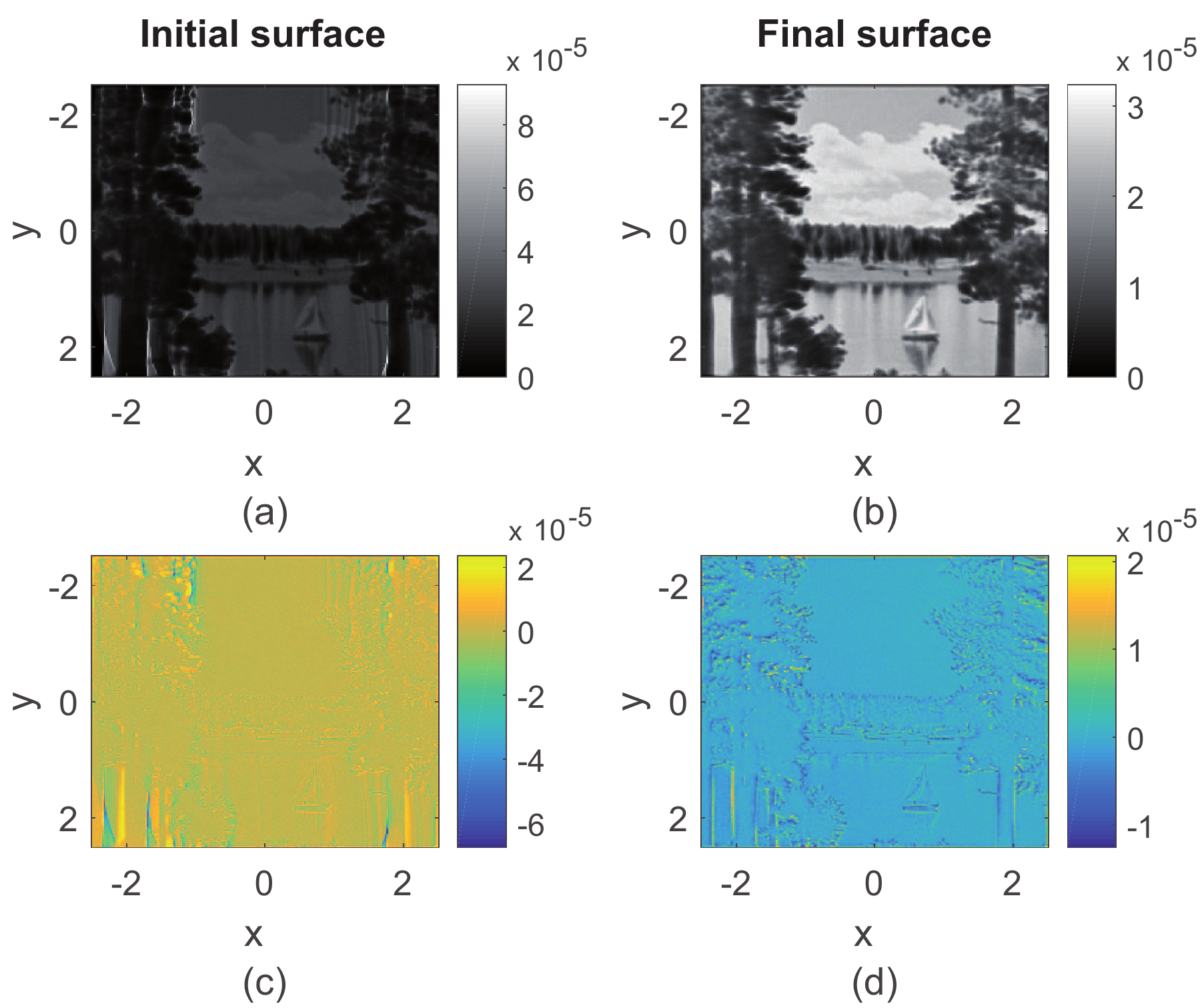}
\end{tabular}
\end{center}
\caption 
{Target irradiance distribution $I_T(x,y)$ from the ray tracing simulation produced by the (a) initial surface and (b) optimized surface. Difference between the predefined distribution $I_T(x,y)$ and the ray tracing simulation $I_{T,RT}(x,y)$ for (c) the initial surface and (d) the optimized surface.}
\label{fig:7}
\end{figure}

As described in section \ref{sec:5}, the function $z_S (x,y)=z(x_S, y_S)$ is interpolated and extrapolated onto a square, uniform grid by \textit{RegularizeData3D()} to give $z(x,y)$. For a fine boundary interpolation, we choose an interpolation grid of $500 \times 500$ points, whereby the minimum and maximum $x$ and $y$ values correspond to the minimum and maximum $x_S$ and $y_S$ values (see Fig. \ref{fig:6}(b)). The function values are then imported to a ray-tracing software. In Fig. \ref{fig:7} and Table \ref{table:2} the results of the ray tracing are shown. From Fig. \ref{fig:7}(a) and \ref{fig:7}(c) it can be seen that the quadratic cost function in Eq. (\ref{eq:12}) provides a reasonable initial iterate for the root-finding process, as argued in section \ref{sec:4}. Figure \ref{fig:7}(b) and \ref{fig:7}(c) and Table \ref{table:2} show results, which are comparable to the example from the previous subsection , except for the slightly lowered efficiency $\eta$ due to the interpolation of $z_S (x,y)$ at the surface boundary.

\begin{table}[!htb]
\centering
\caption{\bf Comparison of $\Delta I_T$ for example ``lake" with a point source.}
\begin{tabular}{ccc}
\hline
	 & $z^{\infty}(x,y)$ & $z(x,y)$ \\
\hline
$rms_{\Delta I_T}$	& $4.1373\cdot 10^{-6}$	& $2.1480\cdot 10^{-6}$	\\
$pv_{\Delta I_T}$	& $9.0955\cdot 10^{-5}$	& $3.1909\cdot 10^{-5}$	\\
$corr_{I_T}$		& $0.8889$				& $0.9668$				\\
$\eta$				& $99.62 \% $			& $99.55 \% $		\\
\hline
\end{tabular}
  \label{table:2}
\end{table}

From the viewpoint of manufacturing, it is usually preferred to have a plane  instead of a spherical surface as the entrance surface of the lens. In the following section it is demonstrated that the design algorithm can be applied directly to these configurations.

\subsection{Astigmatic wavefront}
\label{sec:6.3}

\begin{figure}[!htb]
\begin{center}
\begin{tabular}{c}
\includegraphics[width=\linewidth]{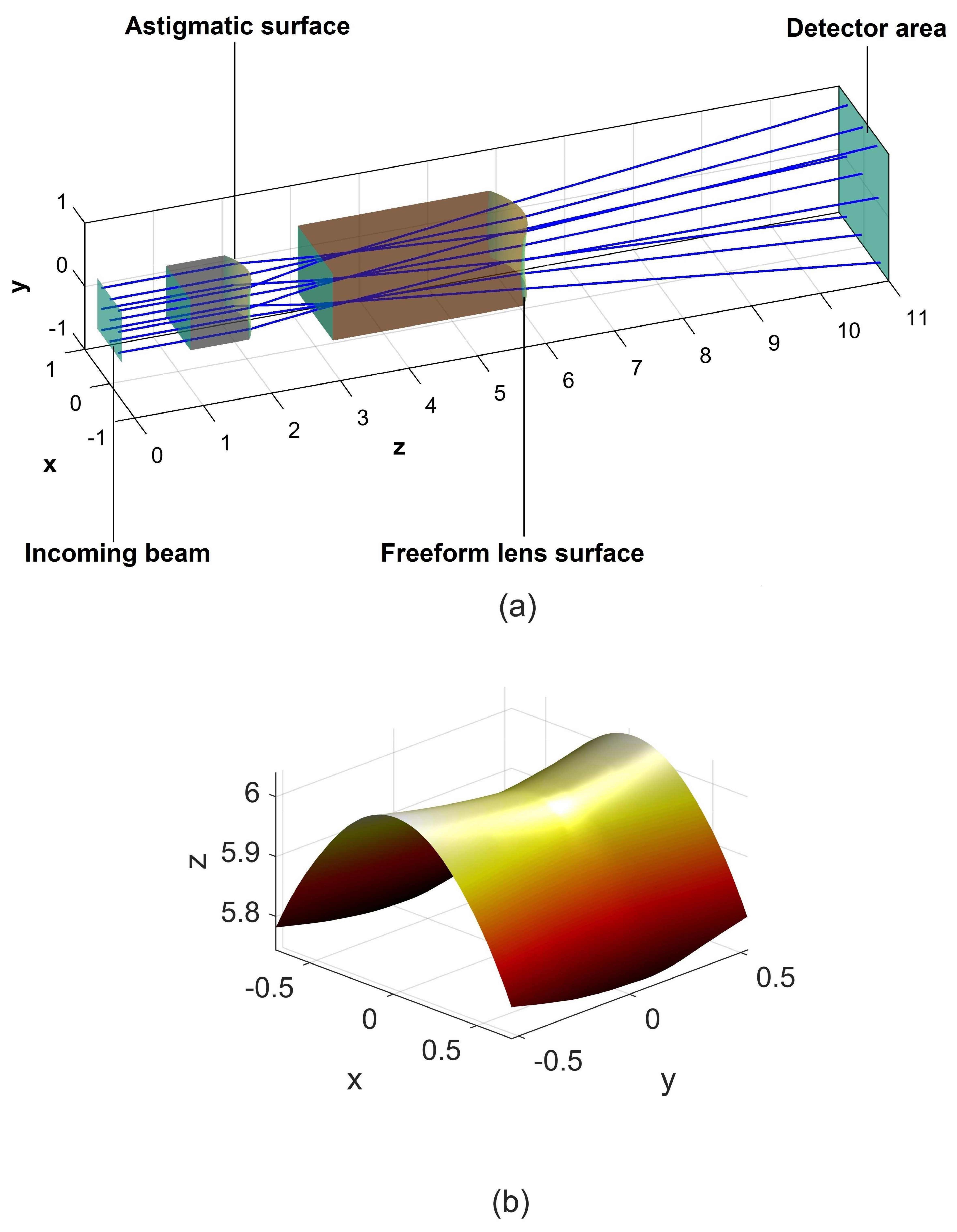}
\end{tabular}
\end{center}
\caption 
{(a) System Layout. The incoming collimated gaussian beam is refracted by a predefined surface to give an astigmatic input wavefront. After another refraction by the plane entrance surface the rays are refracted by the calculated freeform surface to give the required target distribution. (b) Freeform lens surface.}
\label{fig:8}
\end{figure}

\begin{figure}[!htb]
\begin{center}
\begin{tabular}{c}
\includegraphics[width=\linewidth]{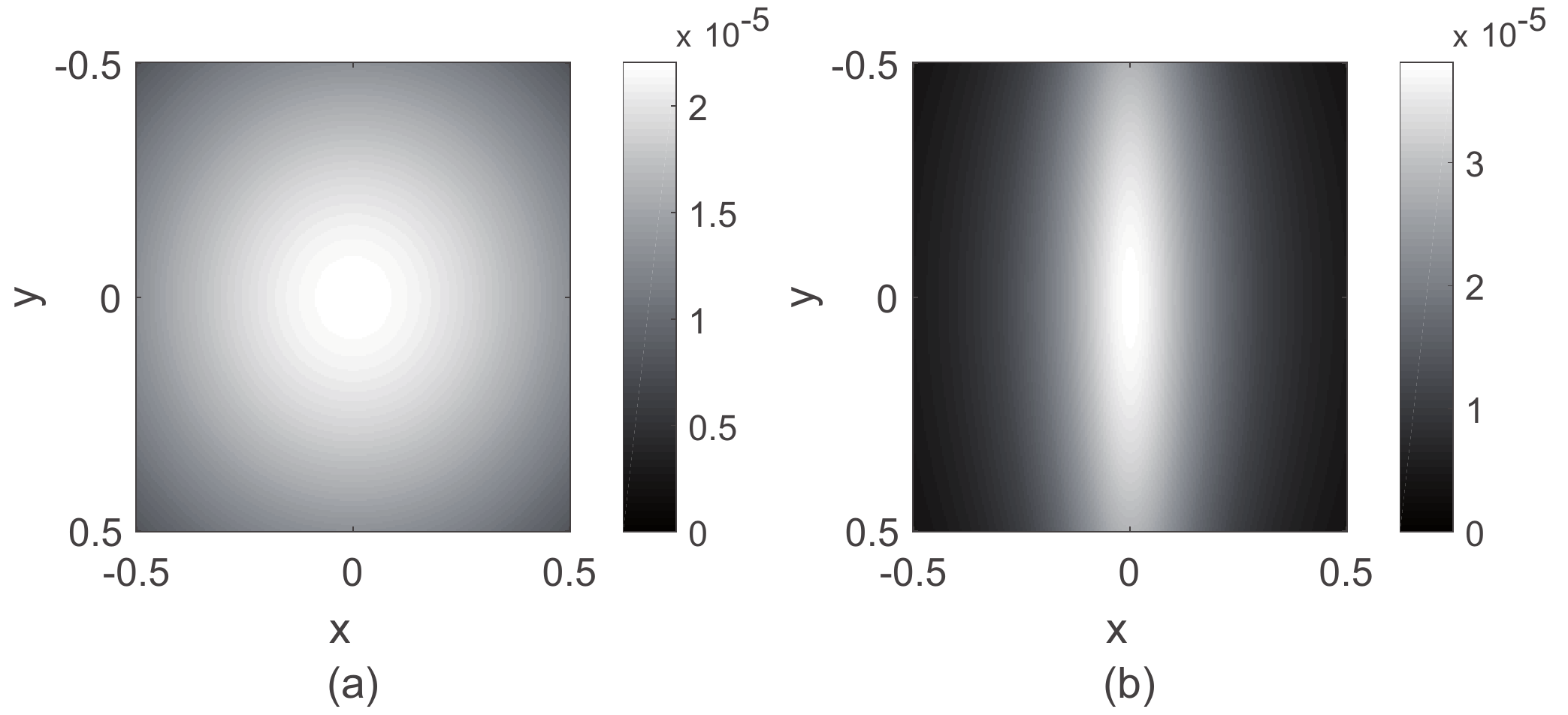}
\end{tabular}
\end{center}
\caption 
{(a) Input gaussian beam before the refraction by the predefined astigmatic surface. (b) Source irradiance distribution $I_S (x,y)$ at $z=5$ (Fig. \ref{fig:8}(a)) after refraction by the plane surface. $I_S (x,y)$ is used as the input distribution for the calculation of $\mathbf{u}^{\infty}(x,y)$ with optimal mass transport and together with the input ray direction vector field $\mathbf{\hat{s}_1}$ and $I_T(x,y)$ as the input of the design algorithm.}
\label{fig:9}
\end{figure}

In this section, we want to explore an example with a nonstandard input vector field $\mathbf{\hat{s}}_1$. We therefore \textit{predefine} a surface $z_{pre}(x,y)=\frac{1}{9} y^2 - \frac{2}{3} x^2$, deforming a collimated beam with a gaussian distribution (Fig. \ref{fig:9}(a)) into a beam with an astigmatic wavefront. The corresponding system layout can be seen in Fig. \ref{fig:8}.
  
As in the previous examples, we need the input vector field $\mathbf{\hat{s}}_1$ and the irradiance distribution on an intermediate plane $z=z_0$. This plane is chosen to be $z_0 =5$ and therefore after the refraction of the astigmatic beam by the plane entrance surface ($z=3$ in Fig. \ref{fig:8}(a)), and the aperture at $z=z_0$ is a square with a side length of $1$. The vector field $\mathbf{\hat{s}}_1$ and the source distribution $I_S (x,y)$ on $z=z_0$ can then be calculated numerically, e.g. from the law of refraction and the Jacobian equation (\ref{eq:2}). $I_S (x,y)$ is shown in Fig. \ref{fig:9}(b) and serves as the input for the calculation of $\mathbf{u}^{\infty}(x,y)$ ($417 sec$).

\begin{figure}[!htb]
\begin{center}
\begin{tabular}{c}
\includegraphics[width=\linewidth]{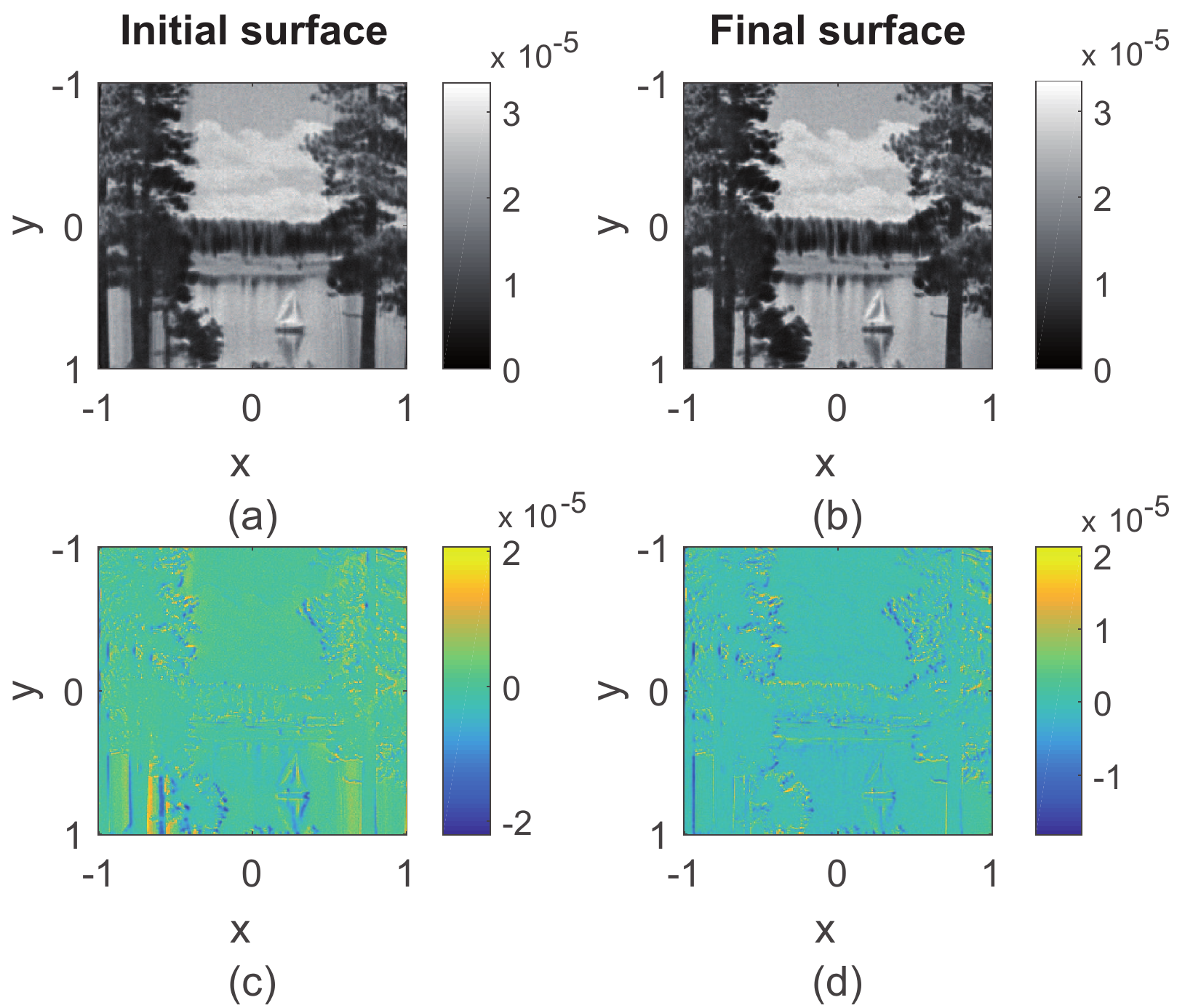}
\end{tabular}
\end{center}
\caption 
{Target irradiance distribution $I_T(x,y)$ from the ray tracing simulation produced by the (a) initial surface and (b) optimized surface. Difference between the predefined distribution $I_T(x,y)$ and the ray tracing simulation $I_{T,RT}(x,y)$ for (c) the initial surface and (d) the optimized surface.}
\label{fig:10}
\end{figure}

For the calculation of the intial surface with Eq. (\ref{eq:15}), we have to fix one point of the freeform surface relative to the plane $z=z_0$, which we choose to be $z(0,0)=z_0 +1$ and the side length of the target area is chosen to be $2$. The calculation of $z_S^{\infty} (x,y)$ took $145 sec$ and the root finding of Eq. (\ref{eq:10}) about $215 sec$. Since the input vector field is noncollimated, the surface $z(x,y)$ is again calculated on a scattered grid and has to be interpolated for the ray-tracing simulation. The results of the ray tracing are shown in Fig. \ref{fig:10} and Table \ref{table:3}. Also for this example, the results show that the initial map is appropriate for the root finding of Eq. (\ref{eq:10}) and that by the presented design algorithm, single freeform surfaces for complex target irradiances and nonplanar or nonspherical input wavefronts can be calculated.

The quantitative differences between the ray-tracing results for the examples are thereby mainly due to interpolation and extrapolation effects. It is also a priori not clear, if for the particular predefined system parameters of each example an exact solution to Eq. (\ref{eq:10}) exists. This might also lead to a degradation of the qualitiy of the ray-tracing results, since an accurate root-finding of the respective nonlinear equation systems might not be possible. In contrast to the first two examples, additional errors are introduced in the third design example by the numerical calculation of $I_S (x,y)$ and the input direction vector field.

\begin{table}[!htb]
\centering
\caption{\bf Comparison of $\Delta I_T$ for example ``lake" with an astigmatic input wavefront}
\begin{tabular}{ccc}
\hline
	 & $z^{\infty}(x,y)$ & $z(x,y)$ \\
\hline
$rms_{\Delta I_T}$	& $3.2072\cdot 10^{-6}$		& $2.7386\cdot 10^{-6}$	\\
$pv_{\Delta I_T}$	& $4.2670\cdot 10^{-5}$		& $3.9421\cdot 10^{-5}$	\\
$corr_{I_T}$		& $0.9252$					& $0.9464$				\\
$\eta$				& $99.23\% $				& $99.20 \% $		\\
\hline
\end{tabular}
  \label{table:3}
\end{table}

\section{Conclusion}

A single freeform surface design algorithm for prescribed input wavefronts and complex irradiance distributions was introduced. This was done by deriving the PDE system (\ref{eq:10}) and Monge-Amp\`ere equation (\ref{eq:11}), which governs the single freeform design process for ideal sources in the geometrical optics limit, from the energy conservation and ray tracing equations in cartesian coordinates. After introducing a possible way to calculate an appropriate initial iterate in section \ref{sec:4}, the PDE system is then solved for the unknown functions $\mathbf{u}(x,y)$ and $z(x,y)$ by a finite difference discretization and a root finding of the resulting nonlinear equation system. The design examples in section \ref{sec:6} show that the presented design algorithm extends the results of Ref. \cite{Boe16_1} for collimated beam shaping to the nonparaxial regime and that the algorithm is applicable to point sources and more general input wavefronts. This might be helpful in applications in which one has to deal with the shaping of e.g. astigmatic wavefronts produced by laser diodes or simplify manufacturing processes. Also the relaxation of physical situations, e.g. for point sources with large opening angles, might be reached by a predefined entrance surface of a freeform lens.

Despite the presented results, there are still interesting problems left for future research and improvements, respectively. For example, a mathematical rigorous justification of the initial iterate construction method presented in section \ref{sec:4} is of interest, since there is no guarantee that the proposed method will always give a suitable intial iterate, despite the arguments given in section \ref{sec:4}. Also the numerical extension of the presented algorithm to more complicated boundaries of the irradiance distributions would help to increase the number of possible applications or energy efficiency, respectively. This requires the numerical implementation of the transport boundary conditions, which are realized for irradiance distributions with square boundary shapes by Eq. (\ref{eq:19}), for more complex boundary shapes and will be investigated in the future. As mentioned in section \ref{sec:3}, a numerical investigation of the Monge-Amp\`ere equation in Eq. (\ref{eq:11}) is of interest. Despite the listed additional complications, the reduced number of design variables can benefit the speed of the root-finding process. Another interesting future goal is the generalization of the presented design algorithm to wavefront and irradiance control with double freeform surfaces by combining it with the results presented in Ref. \cite{Boe17_1} and the investigation of the manufacturability of the calculated freeform surfaces.

\section*{Funding}

We acknowledge support from the German Federal Ministry of Education and Research (BMBF) under contract number 031PT609X within the project KoSimO.

\section*{Acknowledgments}

The authors want to thank Norman Worku for providing and helping them with the MatLightTracer and Johannes Stock for comments on the manuscript.

\end{document}